\documentclass{article}

\usepackage{arxiv}

\usepackage[utf8]{inputenc} 
\usepackage[T1]{fontenc}    
\usepackage{hyperref}       
\usepackage{url}            
\usepackage{booktabs}       
\usepackage{amsfonts}       
\usepackage{nicefrac}       
\usepackage{microtype}      
\usepackage{graphicx}
\usepackage{doi}
\usepackage{multirow}

\title{A selective review of recent developments in spatially variable gene detection for spatial transcriptomics}

\author{ \href{}{\hspace{1mm}Sikta Das Adhikari$^{1,2}$}
        ,
	\href{}{\hspace{1mm}Jiaxin Yang$^2$}
        ,
        \href{}{\hspace{1mm}Jianrong Wang$^{2*}$}
        ,
        \href{}{\hspace{1mm}Yuehua Cui$^{1}$}\thanks{To whom correspondence should be addressed.
        Email: cuiy@msu.edu, wangj164@msu.edu.\\
        $^{1}$Department of Statistics and Probability, Michigan State University, East Lansing, MI 48824, USA.
        and
        $^{2}$Department of Computational Mathematics, Science and Engineering, Michigan State University, East Lansing, MI 48824, USA.}
 \\
}

\date{}

\renewcommand{\headeright}{}
\renewcommand{\undertitle}{}
\renewcommand{\shorttitle}{}

\hypersetup{
pdftitle={A template for the arxiv style},
pdfsubject={q-bio.NC, q-bio.QM},
pdfauthor={David S.~Hippocampus, Elias D.~Striatum},
pdfkeywords={First keyword, Second keyword, More},
}

\begin{document}
\maketitle
\begin{abstract}
\emph{With the emergence of advanced spatial transcriptomic technologies, there has been a surge in research papers dedicated to analyzing spatial transcriptomics data, resulting in significant contributions to our understanding of biology. The initial stage of downstream analysis of spatial transcriptomic data has centered on identifying spatially variable genes (SVGs) or genes expressed with specific spatial patterns across the tissue. SVG detection is an important task since many downstream analyses depend on these selected SVGs. Over the past few years, a plethora of new methods have been proposed for the detection of SVGs, accompanied by numerous innovative concepts and discussions. This article provides a selective review of methods and their practical implementations, offering valuable insights into the current literature in this field.}
\end{abstract}


\section{Introduction}
Recent advancements in Spatially-resolved transcriptomics (SRT) technology have provided comprehensive gene expression data for thousands of genes across multiple samples or spatial spots, accompanied by their respective spatial coordinates. Depending on the specific technology utilized, a sample could represent a single cell (as in the case of STARmap technology), a cell-sized local region (as with HDST technology\cite{HDST}), or a localized region comprising dozens of cells (as seen in Slide-seq\cite{Slide_seq1,SlideSeq2} and Visium technologies). The latest SRT platforms, such as 10x Genomics Visium and Slide-seqV2, encompass thousands of spatial locations within each tissue sample, with future developments poised to achieve even higher resolutions. As technology progresses, the demand for more robust statistical frameworks to effectively analyze spatial data intensifies.\\~\\
Although spatial transcriptomic (ST) data permit addressing a range of distinct questions, a fundamental initial step in the downstream analysis of spatial data is the identification of spatially variable genes (SVGs). These are genes that exhibit variations in expression levels either across the entire tissue or within predefined spatial domains. These genes can potentially unveil tissue heterogeneity and the underlying structural factors that drive distinct expression patterns across spatial locations, thus offering valuable insights into biology.\\~\\
Numerous methods have been developed for the identification of SVGs. These methods encompass a spectrum of approaches, including the utilization of standard spatial statistics measures like Moran's I statistic\cite{moransI} and Geary's C statistic\cite{gearyC} to rank genes based on their spatial autocorrelation. More advanced methods employ model-based approaches such as SpatialDE\cite{SpatialDE}, SpatialDE2\cite{SpatialDE2}, SPARK and its extensions\cite{SPARK}, nnSVG\cite{nnSVG}, BOOST-GP\cite{BOOST-GP}, marked point process frameworks like Trendsceek\cite{Trendsceek} and scGCO\cite{scGCO}, or model-free frameworks like sepal\cite{sepal} and GLISS\cite{GLISS}. Additionally, there are toolboxes, such as MERINGUE\cite{MERINGUE}, Giotto\cite{giotto}, Seurat\cite{Seurat} that integrate some of these methods into comprehensive end-to-end analysis frameworks.\\~\\
Downstream analysis involving SVGs encompasses various tasks, such as spatial clustering, deciphering spatial domains, and identifying spatial domain-specific SVGs. Additionally, there are numerous other downstream analyses that leverage additional information like scRNASeq data, histological images, and more, for tasks such as spatial decomposition of spots, gene imputation, the inference of cell-cell and gene-gene interactions and spatial location reconstruction for scRNA‑seq data. However, this review primarily concentrates on SVG detection frameworks and does not delve into these other downstream analyses.\\~\\
Thus, it is the primary focus of this paper to discuss selected frameworks for SVG identification, serving as a valuable resource for researchers new to this field, enabling them to become acquainted with existing SVG identification frameworks, their unique characteristics, novelty, as well as their pros and cons.
\section{Overview of the frameworks for detecting SVGs }
Generally, in a spatial transcriptomics setup, the available spatial dataset contains gene expression measures/counts for $m$ genes distributed across $N$ known spatial coordinates or spots. This section establishes the key symbols that will be frequently utilized. Specifically, $y=(y_1, y_2, ..., y_N)$ is defined as the gene expression profiles/counts for a given gene across spatial coordinates (referred to as samples or spots), denoted by $s = (s_1, ..., s_N)$. The coordinates of the spatial locations are typically two-dimensional, i.e., $s_i = (s_{i1}, s_{i2})$, but any dimensional coordinates can be employed. The primary objective of these SVG detection models is to ascertain which genes, out of the $m$ genes, are spatially variable across the tissue. In other words, the main goal is to determine whether the gene expression measure $y$ depends on or relates to the spatial locations where the gene expression measures are collected. 

\subsection{Gene expression data and pre-processing step}
The gene expression measure $y$ are generally of count data type from sequencing reads. Various SVG detection models have been developed to specifically model count data following some mandatory filtering and quality control steps. Some examples of these models include SPARK-X\cite{SPARK-X}, BOOST-GP\cite{BOOST-GP}, SINFONIA\cite{SINFONIA}, and GPcounts\cite{GPcounts}. The gene expression count data often exhibit over-dispersion and contain numerous zero values, mainly due to the technology employed for data generation or simply because many genes are poorly expressed for biological reasons. These particular issues in count data are generally taken care of by using negative binomial models which handle over-dispersion well. For the issue of zero-inflation, Zhao et al, 2022 \cite{Zhao2022} showed that modeling zero inflation is not necessary in spatial transcriptomics, thus is not a concern in many method development. On the other hand, some methods, for example SpatialDE\cite{SpatialDE}, nnSVG\cite{nnSVG}, and BOOST-MI\cite{BOOST-MI}, use normalized gene expression data in the model for easy implementation, where in most of cases, the data is modeled using multivariate normal distribution after transformation. Authors in SPARK\cite{SPARK} proposed two different data models, SPARK and SPARK-G which uses count data and normalized data, respectively. The data normalization method is not unique for these methods. The normalization step generally removes the bias due to differences in sequencing depth using size factors and normalizes the data using log transformation(log10 or log2 transformations after adding a pseudo-count value $c$, preferably 1). The method sepal\cite{sepal} uses a slightly different normalization procedure which involves mapping the log-transformed values to the interval [0,1] and using a pseudocount 2. Other normalization methods, such as scran, scuttle, and scater R/Bioconductor packages\cite{package_scran_scuttle_scater2,package_scran_scuttle_scater1}, can also be applied. Table \ref{table:1} provides information on some selective methods together with their required input data type and the implemented model:
\begin{table}[h!]
\begin{center}
\caption{A selective list of methods for SVG detection in ST data analysis along with the required input data type and the implemented model.}\label{table:1}
\begin{tabular}{c c c} 
 \hline
 Method  &Input data type  &Data model   \\ [0.5ex] 
 \hline
Trendsceek\cite{Trendsceek} &Normalized  &Marked point processes\\ [1ex] 
 SpatialDE\cite{SpatialDE} &Normalized  &Multivariate Normal  \\ [1ex] 
  SpatialDE2\cite{SpatialDE2} &Count  &Poisson  \\ [1ex] 
 SPARK\cite{SPARK} &Count &Overdispersed poisson  \\[1ex] 
   SPARK-G\cite{SPARK} &Normalized &Multivariate Normal \\[1ex] 
  SPARK-X\cite{SPARK-X} &Count  &Non-parametric model  \\ [1ex] 
  nnSVG\cite{nnSVG} &Normalized  &Multivariate Normal  \\ [1ex] 
   BOOST-GP\cite{BOOST-GP} &Count  &Zero-inflated negative binomial\\ [1ex] 
   SOMDE\cite{SOMDE} &Normalized  &Multivariate Normal  \\ [1ex] 
   sepal\cite{sepal} &Normalized  & Model-free  \\ [1ex] 
  GLISS\cite{GLISS} &Normalized  & Model-free  \\ [1ex] 
  SINFONIA\cite{SINFONIA} &Count  &Model-free  \\ [1ex] 
  BOOST-MI\cite{BOOST-MI} & Normalized  & Modified Ising model  \\ [1ex] 
  scGCO\cite{scGCO} &Normalized  &Marked point process  \\ [1ex] 
  CTSV\cite{CTSV} &Count  &Zero-inflated negative binomial  \\ [1ex] 
  HEARTSVG\cite{HEARTSVG} &Count  &Model-free  \\ [1ex] 
  MULTILAYER\cite{MULTILAYER} &Normalized  &Model-free  \\ [1ex] 
  GPcounts\cite{GPcounts} &Count  &Negative binomial  \\ [1ex] 
 \hline
\end{tabular}
\end{center}
\end{table}
\subsection{Overview of model-based frameworks}
\subsubsection{Gaussian process(GP) regression based and similar models}
The majority of the methods, including some of the state-of-the-art algorithms to detect SVG, are based on Gaussian process (GP) regression models. For example, one of the first published SVG detection methods, SpatialDE\cite{SpatialDE}, models the normalized gene expression $y$ for a given gene using the following multivariate normal model:
\begin{equation}\label{mGP}
p(y|\mu,\sigma_s^2,\delta,R)\sim N(y|\mu1,\sigma_s^2K+\delta I),
\end{equation}
where the covariance term is decomposed into a spatial and a non-spatial part, where $\delta I$ represents the non-spatial part and $\sigma_s^2K$ is the spatial covariance matrix, whose $(i,j)^{th}$ element in the kernel matrix $K$ denotes the spatial similarity between the $i^{th}$ and $j^{th}$ spot calculated based on the corresponding coordinates $s_i$ and $s_j$. The choice of the kernel function plays a very important role in detecting the spatial correlation present in the gene expressions. More discussion about kernel function is provided in the next subsection. \\~\\
Other methods like SPARK-G\cite{SPARK} (the Gaussian version of SPARK), nnSVG\cite{nnSVG}, and SOMDE\cite{SOMDE} implement similar GP models for modeling normalized gene expression data with some extra features or added level of complexity. SPARK-G and nnSVG provide the option to include extra covariate terms in the model. The covariates or the explanatory variables could contain batch information, cell-cycle information, or other information that is important to adjust for during the analysis. SOMDE is a two-step procedure. This approach involves first utilizing a self-organizing map to cluster neighboring cells into nodes. Subsequently, it employs a Gaussian process to model and analyze the spatial gene expression patterns at the node level.\\~\\
Table \ref{table:1} shows that methods like SPARK\cite{SPARK}, SpatialDE2\cite{SpatialDE2}, BOOST-GP\cite{BOOST-GP}, CTSV\cite{CTSV}, and GPcounts\cite{GPcounts} model count data directly. SPARK models the count data using an overdispersed poisson model where the logarithm of the unknown Poisson rate parameter is assumed to follow a stationary Gaussian process with similar spatial and non-spatial covariance components. BOOST-GP presents a novel Bayesian hierarchical model to analyze spatial transcriptomic data, which models the count data using a zero-inflated negative binomial(ZINB) model. The logarithm of the normalized expression level, which is included in the expectation term in NB, can be seen as a GP with a spatial covariance term representing the spatial variability in case there is a spatial pattern. GPcounts also uses negative binomial distribution to model the UMI(Unique Molecular Identifier) data. SpatialDE2 employs a Generalized Linear Mixed Model (GLMM) for count data modeling. In contrast to GP-based techniques that typically separate covariance into a spatial and a non-spatial component, SpatialDE2 dissects the covariance into several spatial components along with a non-spatial random component. CTSV implements a slightly different technique and does not use the GP model. In CTSV, the gene specific, spot specific and cell-type specific relative mean expression level in the ZINB model is a linear combination of $h_1(s_{.1})$ and $h_2(s_{.2})$ where the functions $h_1(\cdot)$ and $h_2(\cdot)$ represents the underlying true spatial effect modeled with the kernel function in GP model. 
\subsubsection{Statistical inference and selecting kernel function in GP-based frameworks}
Typically, when evaluating the existence of spatial patterns within the data, an assessment is made by testing the alternative hypothesis, which suggests the presence of a spatial covariance term in the model, against the null hypothesis, where the spatial covariance term is set to zero, indicating the absence of spatial variability. This comparison between the model fitted under the alternative hypothesis and the null model forms the basis of a significance testing procedure. This often involves conducting significance tests and drawing conclusions based on p-values in frequentist approaches. For example, in model (\ref{mGP}), testing SVG is equivalent to testing $H_0: \sigma_s^2=0$.\\~\\ 
As previously mentioned, selecting the appropriate kernel function for computing the spatial covariance matrix is a critically important step in identifying spatial patterns within the data. Ideally, the kernel function should accurately capture the true underlying relationship between the $y$ values and the spatial coordinates $s$. In practice, the actual underlying function remains unknown, and the closer the chosen kernel function approximates the true functional form, the more precise the model specification becomes, rendering the test more robust and powerful.\\~\\
SpatialDE employs a squared exponential covariance function(a.k.a Gaussian kernel function or radial basis kernel function) to compute the spatial covariance matrix: 
\begin{center}
    $K_{i,j}=k(s_i,s_j)=exp(-\frac{|s_i-s_j|^2}{2l^2})$
\end{center}
The hyperparameter $l$, recognized as the characteristic length scale or bandwidth, determines how rapidly the covariance decays as a function of distance and is typically chosen by grid search. SOMDE also uses the squared exponential (Gaussian) kernel in their model with ten different length scales and chooses the one that achieves the highest log-likelihood ratio value. GPcounts uses linear or periodic kernel based on BIC values. SPARK asserts that relying on a single kernel restricts the ability to robustly identify spatially variable genes across diverse spatial patterns. Therefore, SPARK (and SPARK-G) adopts an approach involving a total of ten distinct spatial kernels. These comprise five periodic kernels (e.g., Cosine kernels) with varying periodicity parameters and five Gaussian kernels with different smoothness parameters. SPARK proceeds to compute ten p-values, each derived from a different test employing these various kernel functions. These p-values are subsequently combined using the Cauchy combination rule \cite{cauchyrule1,cauchyrule2}. Similar to SPARK, SpatialDE2 incorporates a variety of pre-defined kernels with varying structures and length scales. It also offers the flexibility to conduct an omnibus test as an alternative to independently testing each kernel and subsequently merging the p-values. nnSVG posits that genes can potentially display a vast spectrum of spatial patterns, and using the same set of kernel functions for all of the genes might lead to less powerful tests. They consider the use of an exponential covariance function as a kernel function where the length scale parameter of the kernel function is fitted for each gene, which allows capturing the unique spatial variability pattern of the gene. CTSV uses five different sets of functional forms for $h_1(s_{.1})$ and $h_2(s_{.2})$, which includes linear functions, squared exponential functions, and periodic functions with different sets of scaling parameters and the five p-values calculated from five different forms are combined using the Cauchy combination method. \\~\\
Although different models discussed here have some similarities in testing procedures, the model fitting techniques implemented and the testing procedures utilized are different and are summarized in Table \ref{table:2}.\\
\begin{table}[ht!]
\begin{center}
\caption{List of some popular SVG detection methods with model-fitting and testing information.}\label{table:2}
\begin{tabular}{c c ccc} 
 \hline
  \multirow{2}{*}{Method}  &Bayesian/ &Model fitting and  &Hypothesis testing   \\ 
 &Frequentist&parameter estimation&method\\ [0.5ex]
 \hline
 SpatialDE &Frequentist &Maximizing marginal   &LogLikelihood ratio test\\ 
 &&log likelihood&\\[1ex] 
 SpatialDE2 &Frequentist &Only null model parameters needs   &Score test based\\ 
 &&to be estimated by BLUP& on Zhang and Lin\cite{score_spatialDE2}\\[1ex] 
 SPARK &Frequentist &Approximate-inference algorithm   &Satterthwaite method \\
 &&based on the PQL approach &on the basis of score statistics\\[1ex] 
  SPARK-G &Frequentist &Maximum likelihood &Score test \\[1ex] 
 nnSVG &Frequentist  &Fast optimization algorithms for  & Likelihood ratio test\\ 
 &&NNGP models (BRISC R package)&\\[1ex] 
 BOOST-GP &Bayesian  &Sampling from posterior &Bayes Factor or posterior\\ 
 &&using MCMC &probabilities of inclusion (PPI)\\[1ex] 
  SOMDE &Frequentist  &Gradient optimization  &Likelihood ratio test\\ [1ex] 
 CTSV &Frequentist  &Approx. maximum likelihood  &Wald tests (R \\ 
 && using conjugate gradient(CG) &package pscl)\\
 &&algorithm & \\[1ex] 
 GPcounts &Frequentist &Optimization of log marginal   & Likelihood ratio test\\ 
 &&likelihood by variational &\\
 &&approximation&\\[0.5ex] 
 \hline
 \end{tabular}
\end{center}
\end{table}\\
The statistical power of GP-based methods hinges on the selection of kernel functions\cite{SPARK}, which can complicate the model selection and limit SVG detection power. To address this challenge, the authors in \cite{BOOST-MI} introduced BOOST-MI. This novel approach employs Bayesian modeling of spatial transcriptomics data via a modified Ising model to identify SV genes. As an initial step, BOOST-MI takes normalized gene expression data as input and dichotomizes the normalized expression levels into a binary spatial pattern. Subsequently, BOOST-MI proceeds to identify a wide spectrum of spatial patterns displayed by the genes by inferring the Ising model interaction parameter within a Bayesian framework. It achieves this by generating samples from the posterior distribution of the parameters through a double Metropolis-Hastings (DMH) algorithm\cite{DMH}. Subsequently, it computes the Bayes factor based on these posterior samples, which are then used for selecting SV genes. \\~\\
Trendsceek\cite{Trendsceek}, one of the earliest published SVG detection methods, models data as marked point processes, where they assign points to represent the spatial locations of spots and marks on each point to represent expression levels. The pivotal objective of Trendsceek revolves around evaluating the dependency between the spatial distribution of points and their respective marks through pairwise analyses as a function of the inter-point distances. The underlying premise is that if there exists no dependency between marks and point locations, the resulting scores should remain constant across various distances. A resampling procedure is executed to gauge the significance of a gene’s spatial expression pattern, involving permutations of expression values that create a null model with no spatial expression dependency.\\~\\ 
Similar to Trendsceek, ScGCO\cite{scGCO}(single-cell graph cuts optimization) method also models gene expression data as a marked point process where points represent the spatial locations of measured cells or spots, and marks are discrete gene expression states (such as, down-regulated or up-regulated) associated with points. It analyzes the dependency of points with a specific mark on spatial locations using a hypothesis test. Under the null hypothesis (i.e., no spatial dependency), it assumes that points with a specific mark in a 2D space are distributed in a completely random fashion and can be described by a homogeneous spatial Poisson process. Genes with spatial regions whose number of cells/spots of specific marks are associated with statistically significant low probabilities under the null model are selected as SVG. 
\subsection{Overview of model-free frameworks}
There are other SVG detection methods such as SPARK-X\cite{SPARK-X}, sepal\cite{sepal}, GLISS\cite{GLISS}, and SINFONIA\cite{SINFONIA} which do not attempt to model the data generation process or rely on distributional assumptions. Instead, they use model-free techniques to detect SVGs. The authors introduced sepal\cite{sepal} (Spatial Expression Pattern Locator), an innovative method that leverages transcript diffusion simulations to identify genes exhibiting spatial patterns. It simulates transcript diffusion within the spatial domain and measures the time required for convergence. The core idea is that transcripts with random spatial distributions will converge more quickly or reach a homogeneous state faster compared to those with distinct spatial patterns. Consequently, the diffusion time serves as an indicator of a gene’s degree of spatial variability. Genes with longer diffusion times exhibit less spatial randomness. Therefore, ranking genes based on this indicator and selecting the top-ranked genes as SVGs is a logical approach.\\~\\
SINFONIA\cite{SINFONIA} offers a scalable approach to initially identify spatially variable genes through ensemble strategies as part of its spatial transcriptomic data analysis, with the ultimate goal of deciphering spatial domains. SINFONIA initially identifies the $k$ nearest neighbors in Euclidean space for each spot and builds a Spatial Neighbor Graph (SNG) using the weight matrix where the ($i,j$)th element is determined by a function of the distance between the $i$th and $j$th spot. Next, SINFONIA calculates Moran’s I and Geary’s C statistics for each gene based on the weight matrix $W$ to assess spatial autocorrelation. The underlying concept is that genes with more pronounced spatial autocorrelation exhibit more organized spatial expression patterns.\\~\\
HEARTSVG\cite{HEARTSVG} utilizes a unique, distribution-free, test-based approach that focuses on identifying non-SVGs first and then infers the presence of SVGs using this information. The process involves assessing serial autocorrelations within the marginal expressions across the global spatial context to pinpoint non-SVGs. This, in turn, enables the automatic recognition of all other genes as SVGs, regardless of their spatial patterns. HEARTSVG asserts its superiority in terms of robustness and computational efficiency by abstaining from assumptions about specific underlying spatial patterns for these variable genes.\\~\\
SPARK-X\cite{SPARK-X} is a nonparametric method grounded in the following insight: if $y$ is independent of $s$, then the spatial distance between two locations $i$ and $j$ would also be unrelated to the gene-expression difference between those two locations. SPARK-X constructs two $N \times N$ projection covariance matrices: (1) The expression covariance matrix based on gene expression levels; and (2) the distance covariance matrix based on all spatial locations. It employs a test statistic derived from the product of these two covariance matrices to evaluate the independence between the gene expression ($y$) and the spatial coordinates ($s$).  In simpler terms, if gene expressions are indeed independent of spatial coordinates, the product of these covariance matrices will yield a small value. Conversely, if gene expressions are dependent on the spatial coordinates, the product of the matrices will yield a large value. \\~\\
Similar to the kernel matrix used in methods like SpatialDE or SPARK, the statistical power of the SPARK-X test inevitably hinges on how the distance covariance matrix is constructed and how well it aligns with the true underlying spatial patterns exhibited by the gene of interest. To ensure robust identification of spatially varying genes across diverse spatial expression patterns, SPARK-X explores various transformations of the spatial coordinates ($s$) and subsequently generates distinct distance covariance matrices. Specifically, the algorithm applies five Gaussian transformations with varying smoothness parameters and five cosine transformations to the spatial coordinates ($s$). This process results in the creation of eleven distinct p-values, corresponding to the ten transformed distance covariance matrices and the original one constructed using the original coordinates. These individual p-values are then combined using the Cauchy combination method.\\~\\ 
MULTILAYER\cite{MULTILAYER} treats spatially transcriptomics data as a raster image and uses digital image strategies to resolve tissue substructures. The basic unit in MULTILAYER is the "gexel", gene expression element analogous to a pixel in a digital image. The gene expression levels per gexel relative to the average gene expression are computed within the tissue. Genes are considered upregulated or downregulated when their normalized read counts per gene are above or below the average behavior, respectively. Differentially expressed genes are ranked based on the number of related gexels, providing a rapid view of genes that are overrepresented on the digital map based on their relative expression. 
\\~\\
GLISS\cite{GLISS} (Graph Laplacian-based Integrative Single-cell Spatial
Analysis) utilizes a graph-based feature learning framework to detect and discover SVGs and recover cell locations in scRNA-seq data by leveraging spatial transcriptomic and scRNA-seq data. The workflow involves multiple steps. First, SV genes are identified from ST data using graph-based feature selection. Next, it determines the cells of interest in the scRNA-seq data based on unsupervised learning methods and
leverage these selected SVGs to discover new SVGs in scRNA-seq data. The final goal of this workflow is to cluster genes based on their spatial patterns. \\~\\
The BPS (Big-Small Patch)\cite{BPS} method, introduced in a recent publication, utilizes a non-parametric model for the identification of spatially variable genes in 2D or 3D spatial transcriptomics data. The approach involves taking normalized spatial transcriptomics data as input. It defines big and small patches for each spatial spot based on neighboring spots with larger or smaller radii, respectively. The method then calculates local means of gene expression for both big and small patches. Following this, it calculates the ratio between the variances of local means for each gene, approximating a log-normal distribution for the distribution of these ratios. Subsequently, a p-value is determined for each gene based on this approximated distribution.
\section{Statistical Inference with Multiple Testing Control}
We have previously discussed both model-based and model-free methods for detecting SVGs. The mathematical models employed for capturing the data generation process and the innovative model-free SVG detection technique have proven valuable for uncovering significant SVGs that offer critical biological insights. However, from a statistical perspective, concerns arise regarding the potential for false discoveries of genes that lack genuine spatial variability. This concern becomes more pronounced when a large number of genes are simultaneously tested across most frameworks. If the false discovery rate or type 1 error is not adequately controlled, it may lead to incorrect conclusions and the selection of numerous genes that exhibit false spatial variability.\\~\\
Various methods have been developed for multiplicity correction (MC) to address this concern. Some methods analytically constrain the false discovery rate (FDR) to remain below a predetermined threshold, while others do not analytically control the FDR and simply select a user-specified number of top genes as SVGs. Researchers may choose a method that aligns better with their research goals and the type of downstream analysis they intend to perform. In Table \ref{table:3}, we present an overview of these methods, organized around these critical questions. The permutation-based method is usually considered as the golden standard method as it is purely data-driven and distribution free. However, it is the least scalable one since it is computationally more demanding. The FDR-based methods have been the commonly applied ones since they offer type I error control while maintaining high power compared to the Bonferroni method. Nevertheless, depending on the downstream analysis goal, it is not necessary to strictly enforce the MC rule. For example, when the goal is to find the low dimensional embedding of genes, such as in spatial PCA analysis \cite{SPatialPCA} 
people usually choose top ranked genes for further analysis. In such cases, strictly enforcing MC is not needed. 
\begin{table}[ht!]
\begin{center}
\caption{List of methods and the procedures used for multiple testing control.}\label{table:3}
\begin{tabular}{cc c} 
 \hline
 \multirow{2}{*}{Method} &If framework analytically   &\multirow{2}{*}{How SVGs are selected} \\ 
 &controls FDR& \\
 \hline
Trendsceek &Yes  & Permutation based p-values,\\ 
&&Benjamini–Hochberg procedure\cite{Benjamini_Hochberg} for MC\\[1ex] 
 SpatialDE &Yes  &Analytically estimated p-values, \\ 
 &&q-value method\cite{Q-value} for MC\\ [1ex] 
 SpatialDE2 &Yes  &Analytically estimated p-values, \\
 &&Benjamini–Yekutieli procedure\cite{Benjamini_Yekutieli} for MC  \\[1ex] 
 SPARK &Yes &Analytically estimated p-values,\\
 &&Benjamini–Yekutieli procedure for MC  \\[1ex] 
  SPARK-G &Yes &same as SPARK  \\[1ex] 
 SPARK-X &Yes & same as SPARK \\ [1ex] 
 nnSVG &Yes  &Analytical approximate p-values, \\
 &&Benjamini–Hochberg method for MC\\ [1ex] 
  BOOST-GP &Yes  &Based on Bayesian FDR controlled PPI threshold\\ [1ex] 
  SOMDE &No  &Top ranked genes based  on \\  
  && spatial variability score\\[1ex]
  sepal &No  &Top $k$ genes with highest ranks   \\ [1ex] 
 GLISS &Yes  &Permutation based p-values,\\
&& Benjamini–Hochberg procedure for MC \\ [1ex] 
 SINFONIA &No  &Top $k$ genes with highest score  \\ 
 && and an ensemble technique\\ [1ex] 
 BOOST-MI &No  &Based on specific Bayes Factor threshold\\ [1ex] 
 scGCO &Yes  &Analytically estimated p-values,\\ 
 &&Benjamini–Hochberg procedure for MC\\[1ex] 
 CTSV &Yes  &Analytically estimated p-values,\\
 &&q-value method for MC\\ [1ex] 
 HEARTSVG &Yes  &Analytically estimated p-values,\\ 
 &&MC by Bonferroni/Holm/Hochberg\\[1ex] 
 MULTILAYER &No  & Based on the two-fold threshold \\
 && of a test statistic\\ [1ex] 
 GPcounts &Yes  &Analytical or permuted p-values,\\  
 &&q-value method for MC\\[1ex]
 \hline
\end{tabular}
\end{center}
\end{table}
\section{Exploring Performance, Advantages, and Limitations}
In the preceding sections, we have explored the complexities associated with spatial count data. In many instances, these count data are characterized by sparsity and overdispersion. Section 2 of this review classifies modeling frameworks based on whether they directly model the count data or opt for modeling the normalized data. Some literatures \cite{SPARK-X, BOOST-GP} argue against modeling normalized data with a Gaussian distribution due to concerns that such a parametric approximation may result in reduced statistical power and difficulties in controlling type 1 errors, especially when dealing with small p-values.\\~\\
On the other hand, methods that employ normalized count data, such as SpatialDE, SPARK-G, and nnSVG, offer advantages, including simpler model structures and reduced computational challenges. Notably, SPARK employs a dual modeling approach, encompassing both an overdispersed Poisson model (SPARK) and a Gaussian model (SPARK-G) for count data analysis. They declare that SPARK-G exhibits significantly improved computational efficiency compared to the Poisson version of SPARK. Moreover, SPARK-G may demonstrate greater resilience to model misspecification, potentially enhancing its effectiveness in specific data applications.\\~\\
Although many researchers prefer to model count data directly, there is no consensus on the preferred approach for directly modeling count data either. While some opt for Poisson distribution models, others argue that it may be insufficient to address issues of overdispersion, suggesting that a negative binomial distribution is more suitable in such cases. Furthermore, when data exhibit extreme sparsity, the utilization of a zero-inflated Poisson or negative binomial model may be more logical, although it tends to introduce greater complexity into the model. But we need to note that direct modeling of sparse count data with a negative binomial distribution or other over-dispersed Poisson distributions incurs algorithm stability issues \cite{SPARK-X,stat_book1, GPcounts}.\\~\\ 
With the continuous evolution of spatial transcriptomic technologies, researchers now have access to increasingly vast and high-resolution spatial datasets. Analyzing these extensive datasets demands the use of efficient and scalable methods for downstream analysis. Notably, approaches like Trendsceek and BOOST-GP impose substantial computational demands. In a study referenced from SRTsim\cite{SRTsim}, it was observed that when applying these methods to synthetic data, Trendsceek (v.1.0.0) required approximately 10 hours, while BOOST-GP needed about 8 hours to analyze a single synthetic dataset containing 1000 genes and 673 locations. In the same research context, SOMDE (v.0.1.8) struggled, failing to process nearly 90 percent of the genes and yielding NA values.\\~\\ 
Another comprehensive comparison, outlined in a review paper\cite{review_benchmark}, assessed the performance of various SVG detection methods. The evaluation considered computational time and memory usage across 20 diverse spatial datasets, each varying in the number of spots or samples. Among the methods examined, including SpatialDE, SPARK-X, nnSVG, SOMDE, Giotto KM, and Giotto Rank (both are implemented in the Giotto package), SPARK-X emerged as the swiftest, with SOMDE following as the second-best option, albeit notably slower than SPARK-X. SpatialDE exhibited poorer performance in larger datasets, while nnSVG proved faster than SpatialDE for larger datasets but relatively slower for datasets with fewer spatial locations. In particular, SPARK-X \cite{SPARK-X} scales linearly with the number of spatial locations, while other methods scale cubically (e.g., SpatialDE and SPARK) or quadratically (SpatialDE2).\\~\\
In terms of peak memory usage, study \cite{review_benchmark} revealed that SOMDE consumed the least memory, with SPARK-X ranking second. Conversely, SpatialDE demonstrated high peak memory consumption. Considering the trade-off between speed and memory usage, SPARK-X and SOMDE emerged as the two most efficient methods, as determined by the experiment. Furthermore, the evaluation included other methods such as Giotto KM, Giotto Rank, and Moran's I , but none of these alternatives matched the efficiency of SPARK-X or SOMDE based on the experimental findings.\\~\\
In summary, each modeling framework comes with its own set of pros and cons, necessitating careful consideration of the trade-off between computational efficiency/cost and performance when selecting the most suitable approach for analyzing spatial count data.
The model-free or nonparametric approaches do not try to capture the data generation process and offer alternative frameworks to detect SVG. Most of the method frameworks are very intuitive but each comes with its own sets of restrictions or assumptions. For example, Trendsceek is a resampling-based method, which incurs a substantial computational load, rendering its application impractical for extensive ST datasets. SPARK-X exhibits impressive performance for high dimensional data, but the authors recommend using it with large sample (e.g., spot) size, say 3,000 or more.
\section{Assessing Input Data and Model Outputs}
For the various methodologies we reivewed so far, some of these approaches primarily focus on identifying genes that exhibit spatial variability across the entire tissue, exemplified by methods like SpatialDE and SPARK. In contrast, others are additionally equipped to detect genes with spatial variability within predefined spatial domains, as seen in nnSVG. Other methods like SpaGCN\cite{SpaGCn} and STAMarker\cite{STAMarker} are designed to identify Spatial domains and detect SVGs within spatial domains.\\~\\
Additionally, certain methods aim to identify SVGs to facilitate downstream analysis. For instance, SINFONIA, as cited in this work, provides a scalable approach for the initial identification of spatially variable genes using ensemble strategies within the context of spatial transcriptomic data analysis. The ultimate objective of this method is to decipher distinct spatial domains within the tissue. \\~\\
Furthermore, some of these methods leverage additional information as input, such as single-cell RNA sequencing (scRNA-seq) data, spatial domain information, or tissue-specific markers, in conjunction with spatial transcriptomic data. For instance, CTSV requires scRNA-seq data and a set of marker genes as input alongside the spatial transcriptomic data. It employs deconvolution techniques like SPOTlight\cite{SPOTlight}, RCTD\cite{RCTD}, or SpatialDWLS\cite{spatialdwls} to estimate cell-type proportions for each spatial spot. Ultimately, this approach identifies spatially variable genes specific to different cell types. Similarly, Trendsceek identifies genes with significant spatial trends and subsequently determines the subset of cells occupying spatial regions of interest.\\~\\
Given the distinct ultimate objectives and input criteria for each method, it would be unfair to evaluate their performance solely based on a single parameter. Rather, the utility or superiority of these frameworks depends on the researcher's specific goals and the nature of their research inquiries. In this context, we present a table that combines these selective frameworks, including details about their typical inputs and primary research objectives (see Table \ref{table:4}).

\begin{table}[ht!]
\begin{center}
\caption{List of selective methods with input data type and main goal.}\label{table:4}
\begin{tabular}{c  c c} 
 \hline
 Method(publication) &Input data &Main goal \\ 
 \hline
 SpatialDE(2018) &ST data &  Finding SVG\\ 
 &&Spatial gene-clustering\\[1ex] 
 SpatialDE2(Archived,2021) &ST data &Tissue region segmentation\\ 
 &&Finding SVG\\
 &&Spatial gene-clustering\\[1ex] 
 Trendsceek(2018) &ST data &Finding SVG\\ 
  &scRNA-Seq data& Identifying cells in spatially \\
 &&significant gene expression regions\\[1ex] 
 SPARK(2020) &ST data &Finding SVG  \\[1ex]
 SPARK-X(2021) &ST data & Finding SVG  \\[1ex]
  nnSVG(2023) &ST data & Finding SVGs across tissue\\
  &&or within spatial domains\\[1ex]
   BOOST-GP(2021) &ST data   &Finding SVG\\ [1ex] 
   SOMDE(2021) &ST data &Finding SVG\\ [1ex] 
 sepal(2021) &ST data  &Finding SVG \\  
 &&Spatial gene-clustering\\[1ex]
   GLISS(Archived,2020)&ST data &Finding SVG, recovering cell\\ 
   &scRNA-seq data  &  locations in scRNA-seq data\\
   &&and gene-clustering\\[1ex] 
  SINFONIA(2023) &ST data  &Finding SVG for\\  
  &&deciphering spatial domains\\[1ex]
  BOOST-MI(2022) &ST data  &Finding SVG\\ [1ex] 
 scGCO(2022)  &ST data &Finding SVG \\[1ex] 
 CTSV(2022) &ST data  &Detecting cell-type-specific \\ 
 &scRNA-seq &SVG\\
 &set of marker genes&\\[1ex] 
  HEARTSVG (Archived,2023) &ST data  &Detecting SVG and spatial domain\\ [1ex] 
  MULTILAYER(2021) &ST data  &Detecting SVG, dimensionality  \\ 
  & &reduction, spatial clustering and more\\[1ex] 
 GPcounts(2021) &ST data  & Finding SVG, identifying gene-specific \\ 
  &scRNA-seq data &branching locations and more\\[1ex] 
 SpaGCN(2021) &ST data  &Identifying spatial domains  \\
   &histology image data &and SVG in domain\\[1ex] 
   STAMarker(Archived,2022)&ST data & Spatial domain-specific variable genes  \\[1ex] 
 \hline
\end{tabular}
\end{center}
\end{table}
\section{Publicly Accessible Code for Major SVG Detection Methods}
Every prominent SVG detection method featured in this paper has made its code publicly available. Certain frameworks have packages published in CRAN or available as Python modules, while others have shared their code on Github, and the package can be installed directly from Github. Here, we have compiled a list of the packages and repositories associated with these techniques, along with the coding language they have used (see Table \ref{table:5}). This compilation aims to facilitate convenient access to their respective code bases, making it easier for researchers to choose a method based on their preferred programming language.
\begin{table}[ht!]
\begin{center}
\caption{List of methods with implementing code language and package site.}\label{table:5}
\begin{tabular}{c c c } 
 \hline
 Method  &Code language & Package or GitHub or vignette \\ 
 \hline
 Trendsceek &R & \url{https://github.com/edsgard/trendsceek} \\ [1ex] 
SpatialDE &Python & \url{https://github.com/Teichlab/SpatialDE} \\ [1ex]
SpatialDE2&Python&\url{https://github.com/PMBio/SpatialDE}\\[1ex] 
 SPARK   &R & \url{https://github.com/xzhoulab/SPARK}   \\
 SPARK-G      & &\url{https://xzhoulab.github.io/SPARK/01_about/} \\
  SPARK-X&&     \\[1ex]
  nnSVG  &R &  \url{https://bioconductor.org/packages/release/bioc/html/nnSVG.html}\\
   && \url{https://github.com/lmweber/nnSVG}\\
   BOOST-GP &R/C++ & \url{https://github.com/Minzhe/BOOST-GP} \\ [1ex] 
   SOMDE &Python& \url{https://pypi.org/project/somde} \\ 
   && \url{https://github.com/XuegongLab/somde}\\ [1ex] 
 sepal &Python& \url{ https://github.com/almaan/sepal} \\ [1ex] 
 & &\url{10.5281/zenodo.4573237}\\
 GLISS & Python & \url{https://github.com/junjiezhujason/gliss} \\ [1ex] 
  SINFONIA &Python & \url{ https://github.com/BioX-NKU/SINFONIA} \\ [1ex] 
  BOOST-MI &R &\url{https://github.com/Xijiang1997/BOOST-MI} \\ [1ex] 
 ScGCO &Python & \url{https://github.com/WangPeng-Lab/scGCO} \\ [1ex] 
CTSV &R & \url{https://bioconductor.org/packages/devel/bioc/html/CTSV.html} \\ [1ex] 
 HEARTSVG &R & \url{https://github.com/cz0316/HEARTSVG.git} \\ [1ex] 
 MULTILAYER &Python & \url{https://github.com/SysFate/MULTILAYER } \\ [1ex] 
 GPcounts &Python &\url{https://github.com/ManchesterBioinference/GPcounts}\\[1ex]
 BPS &Python & \url{https://github.com/juexinwang/BSP/} \\[1ex] 
\hline
\end{tabular}
\end{center}
\end{table}
\section{Concluding Remarks}
We systematically reviewed recently developed frameworks for identifying spatially variable genes and grouped them into different categories and delved into the unique aspects of their models and underlying principles. Here, we provide a brief discussion encompassing various facets, including preprocessing steps, modeling frameworks, inference techniques, scalability, and practical applicability of these frameworks. We explored the performance of select methods as reported in previously published papers. Nevertheless, it is essential to note that we refrained from conducting evaluations based solely on the number of SVGs detected or the trade-off between statistical power and FDR. This decision arises from the fact that the methods discussed in this paper often serve different research objectives, each tailored to specific research questions. For example, a method primarily focused on spatial clustering may yield similar outcomes when considering the top 100 genes versus the top 110 genes. In contrast, a method geared toward accurately identifying genuine SVGs and scrutinizing individual SVGs to glean deeper insights into biological mechanisms may prioritize stringent control of false discovery rates, making it a pivotal concern in their evaluation. The evaluation criteria must align with the unique goals and nuances of each method, akin to comparing apples to oranges when attempting to gauge their performance solely based on the number of SVGs selected.\\~\\ 
This paper\cite{review_benchmark} has previously investigated several methods for detecting spatial gene expression variations and benchmarked their performance based on different measures. It reported that, although each SVG detection method successfully identifies a significant number of SVGs, there is limited overlap in the SVGs detected when a significance cutoff is applied to filter the SVGs. The study's simulation analysis revealed that, in most cases, the estimated FDRs do not accurately reflect the true FDRs. These findings indicate that there is room for improvement in the commonly used methods for SVG detection and their associated FDR control approaches.\\~\\
In the context of Gaussian process based methods, one potential issue could be related to the selection of kernel function. For instance, as an improvement to spatialDE, approaches like SPARK and SPARK-X employ a variety of different kernels to robustly identify various traits. However, they apply the same set of parameter values to all genes, even when these genes may exhibit vastly different spatial patterns. While nnSVG offers improvements by allowing gene-specific kernel function parameter selection, it relies on a single type of kernel function. This opens room for further methodological development for optimal kernel selection when kernel-based methods are applied for SVG detection. \\~\\
Furthermore, model-free techniques, in many cases, do not analytically control FDR, making it challenging to establish a specific cutoff for selecting SVGs. Many methods claim to detect more SVGs than others, often undetected by alternative methods. However, the mere detection of more SVGs does not necessarily indicate the superiority of a framework if it does not effectively control the FDR. If the goal is to pinpoint the top $k$ (say 1000) SVGs for subsequent analysis without the necessity of precisely quantifying detection uncertainty, these methods can be employed. However, for a more rigorous approach, it is crucial to implement stringent FDR control measures to prevent false discoveries. In our empirical analysis, we observed that numerous methods exhibit elevated false positive rates with inflated p-values (data not shown). There is an urgent demand for the development of more rigorous statistical approaches to enhance false positive control.\\~\\
In summary, we have provided a selective survey of recently published and archived literature on SVG detection, offering an analysis of their practical utility, adaptability, innovation, and constraints from various practical perspectives. This effort aims to facilitate new researchers in gaining a holistic understanding of the available methods and assist them in selecting a framework aligned with their specific research needs and questions.
\section*{Acknowledgement}
This work was supported, in part, by awards R01GM131398 from the National Institutes of Health and NSF1942143 from the National Science Foundation.

\bibliographystyle{unsrtnat}






\end{document}